\documentclass[epj,final]{svjour}
\usepackage{latexsym,amssymb,amsmath}
\usepackage{graphics}

\begin{document}

\title{Nonlinear electrical conductivity in a 1D granular medium}
\author{E.~Falcon
\thanks{\email{Eric.Falcon@ens-lyon.fr},
URL http://perso.ens-lyon.fr/eric.falcon/}
\and B.~Castaing
\and M.~Creyssels
}             
\offprints{Eric Falcon}          % Insert a name or remove this line
\institute{Laboratoire de Physique de l'\'Ecole Normale Sup\'erieure de Lyon\\
UMR 5672 - 46 all\'ee d'Italie, 69007 Lyon, France}
\date{Received: date / Revised version: date}
\abstract{We report on observations of the electrical transport within a chain of metallic beads (slightly oxidised) under an applied stress.  A transition from an insulating to a conductive state is observed as the applied current is increased. The voltage-current ($U$--$I$) characteristics are nonlinear and hysteretic, and saturate to a low voltage per contact (0.4 V).  Our 1D experiment allows us to understand phenomena (such as the ``Branly effect'') related to this conduction transition by focusing on the nature of the contacts instead of the structure of the granular network. We show that this transition comes from an electro-thermal coupling in the vicinity of the microcontacts between each bead -- the current flowing through these contact points generates their local heating which leads to an increase of their contact areas, and thus enhances their conduction. This current-induced temperature rise (up to 1050$^{\rm o}$C) results in the microsoldering of the contact points (even for voltages as low as 0.4 V). Based on this self-regulated temperature mechanism, an analytical expression for the nonlinear $U$--$I$ back trajectory is derived, and is found to be in very good agreement with the experiments. In addition, we can determine the microcontact temperature with no adjustable parameters. Finally, the stress dependence of the resistance is found to be strongly non-hertzian due to the presence of the surface films. This dependence cannot be usually distinguished from the one due to the disorder of the granular contact network in 2D or 3D experiments.
}

\maketitle

\section{Introduction}
\label{intro}
Coheration or the ``Branly effect'' is an electrical conduction instability which appears in a slightly oxidized metallic powder under a constraint \cite{Branly90}. The initially high powder resistance falls several orders of magnitude as soon as an electromagnetic wave is produced in its vicinity. Although discovered in 1890 and used for the first wireless radio transmission \cite{Bridgman01}, this instability and other related phenomena of electrical transport in metallic granular media are still not well understood \cite{Falcon03turb}. Several possible processes at the contact scale have been invoked without any clear demonstrations: electrical breakdown of the oxide layers on grains \cite{Kamarinos75,Kamarinos90}, modified tunnel effect through the metal - oxide $\sim$ semiconductor - metal junction \cite{Holm00}, attraction of grains by molecular or electrostatic forces \cite{Gabillard61,Salmer66}, local soldering of micro-contacts by a Joule effect \cite{Vandembroucq97,Dorbolo03bis} also labelled as ``A-fritting'' \cite{Holm00}; each being combined with a global process of percolation \cite{Kamarinos75,Kamarinos90,Gabillard61,Salmer66,Vandembroucq97}.

Understanding the electrical conduction through granular materials is a complicated many body problem which depends on a large number of parameters: global properties concerning the grain assembly (i.e. statistical distribution of shape, size and pressure) and local properties at the contact scale of two grains (i.e. degree of oxidization, surface state, roughness). Among the phenomena proposed to explain the coheration, it is easy to show that some  have only a secondary contribution.  For instance, since coheration has been observed by Branly with a single contact (crossed cylinders or tripod) \cite{Branly02,Branly02bis}, or with a column of beads \cite{Branly99} or disks \cite{Branly98}, percolation can not be the predominant mechanism. Moreover, when a powder sample \cite{Calzecchi84} or just two beads in contact \cite{Guthe00,Guthe01} are connected in series with a battery, a coheration is observed at high enough imposed voltage, in a similar way as the action {\em at distance} of a spark or an electromagnetic wave. In this paper, we deliberately reduce the number of parameters, without loss of generality, by focusing on the electrical transport within a chain of metallic beads directly connected to an electrical source.  As with the acoustical propagation in granular media \cite{Coste97,Gilles03}, such one-dimensional experiments facilitate the understanding of the electrical contact properties, and is a first step toward more realistic media, such as a 2-D array of beads (including disorder of contact) \cite{Falcon03}, and powder samples (including disorder of position). Despite some earlier studies in 1900, with a 2 bead ``coherer'' showing nonlinear characteristics and saturation voltage \cite{Guthe00,Guthe01,Sengupta98,Fisch04,Blanc05,Weiss06}, surprisingly no 1D-work has been attempted to tackle this problem.

The second motivation of our work is to know the pressure dependence of the electrical resistance, $R$, of a granular packing, which also remains an open problem. It was first measured in the case of a contact between two conductors, submitted to a force, $F$, in order to determine the real area of contact \cite{Bowden39}. Indeed, careful attention was paid to initially break any oxide layers at the surface \cite{Bowden39}, or to work with noble metals \cite{Wilson55}, or specific surface coatings \cite{Boyer88} in order to get reproducible results which follow; $R \sim 1/F^{1/3}$, in agreement with the elastic Hertz law. However, when superficial oxide and/or impurity layers are present, this scaling is found again to be a power law but with an exponent greater than 1. This has been observed in a 3D packing of steel beads \cite{Ammi88,Zhuang95} under weak (elastic) compression, or for strongly compressed powders \cite{Euler78}. This anomalous exponent is ascribed either to a superficial contaminant film \cite{Ammi88,Zhuang95}, or a combinaison of a contaminant film and the degree of contact disorder in the packing \cite{Ammi88}. Here again, a 1D experiment should allow to answer if this exponent is driven by contact properties since the effect of contact disorder is absent.

\begin{figure}[h]
\resizebox{1\columnwidth}{!}{%
  \includegraphics{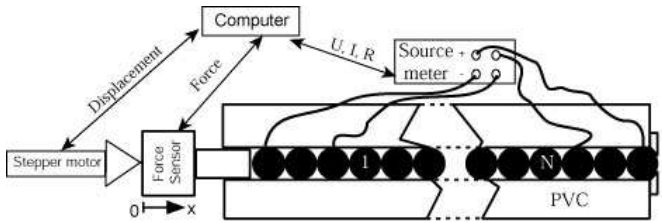}
}
\caption{Schematics of experimental setup}
\label{fig01}  
\end{figure}

\section{Experimental setup}
\label{sec:1}
The experimental setup is sketched in Fig.\ \ref{fig01}. It consists of a chain of 50 identical stainless-steel beads, each 8 mm in diameter, with a tolerance of $4 \mu$m on diameter, and $2 \mu$m on sphericity \cite{Marteau}. The physical properties of the beads are summarized in Table~\ref{properties}. The beads are surrounded by an insulating framework of polyvinylchloride (PVC). It consists of two parts, each one 30 mm high, 40 mm wide and 400 mm long, with a straight channel having a squared section with 8.02 mm sides milled in the lower part to contain the beads. A very small clearance of 2/100 mm is provided in the channel, so that the beads move freely along the chain axis but not in the perpendicular direction. A static force $F$ is applied to the chain of beads by means of a piston (8 mm diameter duralumin cylinder), and is measured with a static force sensor (FGP Instr. 1054)  with a 6.1 mV/N sensitivity in the range from 1 to 500 N. A 1.8 degree stepper motor (RS 440-442) fitted to a gearhead (gear ratio 25:1) is linked to an endless screw, with a 1 mm thread, in order to axially move the piston and the force sensor with a $0.2 \mu$m/step precision. The number of motor steps is measured with a counter (Schlumberger 2721) to determine the piston displacement necessary to reach a specific force. Electrical contacts between the chain and the electrical source are made by soldering leads on particular beads, and the measurements are performed in a four-wire configuration. Note that the lowest resistance of the whole chain (about 3 $\Omega$) is always found much higher than the electrode or the stainless steel bulk material. The bead number $N$ between both electrodes is varied from 1 to 41 by moving the electrode beads within the chain. DC voltage (resp. current) source is supplied to the chain by a source meter (Keithley K2400) which also gives a measure of the current (resp. voltage). The maximum power output is 22 W (210V at 0.105 A or 21V at 1.05A). During a typical experiment, we chose to supply the current ($10^{-6} \leq I \leq 1$ A)  and to simultaneously measure the voltage $V$ and the resistance $R$. The current is supplied during a short time ($\lesssim 1$ s) in order to avoid possible Joule heating of continuous measurements. We note that similar results have been found when repeating experiments with imposing the voltage ($10^{-2} \leq U \leq 2\ 10^{2}$ V) and measuring current and resistance.  The results reported here are highly reproducible.

\begin{table}[hb]
\caption{Relevant mechanical and electrical properties of stainless steel beads used in the chain (norms: AISI 420C, AFNOR Z40C13, grade IV)  \cite{Marteau} or for another stainless steel type (AISI 304)  \cite{CERN}.}
\label{properties}  
\begin{tabular}{lll}
\hline\noalign{\smallskip}
 & Signification & Value  \\
\noalign{\smallskip}\hline\noalign{\smallskip}
$r$ & Bead radius & 4 mm $\pm 2 \mu$m \cite{Marteau}\\
$R_a$ & Roughness & 0.1-0.2 $\mu$m \cite{Marteau}\\
$\rho$ & Density & 7750 kg/m$^3$  \cite{Marteau}\\
$\nu$ & Poisson's ratio & 0.27 \\
$E$ & Young's modulus & 1.95 10$^{11}$ N/m$^2$ \cite{CERN}\\
$\rho_{el}$ & Electrical resistivity 20$^{\rm o}$C & 72 $\mu \Omega$.cm \cite{CERN}\\
 &\qquad  \qquad \qquad \qquad \ \ 650$^{\rm o}$C & 116 $\mu \Omega$.cm \cite{CERN}\\
$\lambda$ & Thermal conductivity 20$^{\rm o}$C & 16.2 W/(Km) \cite{CERN}\\
  &  \qquad \qquad \qquad \qquad \ \ \ \ \ 500$^{\rm o}$C & 21.5 W/(Km) \cite{CERN}\\
$T_{mel}$ & Approx. melting point & 1425 $^{\rm o}$C \cite{CERN}\\
\noalign{\smallskip}\hline
\end{tabular}
\end{table}

\section{Mechanical Behaviour}
\label{meca}
The relation between the force $F$ applied on two identical spheres and the distance of approach $\delta$ of their centers is given by linear elasticity through the so-called Hertz law \cite{Johnson85},
\begin{equation}
F=\frac{E\sqrt{2r}}{3(1-\nu^2)}\delta^{3/2}{\rm \ ,}
\label{hertz}
\end{equation}
and the ``apparent'' radius of the circular contact by \cite{Johnson85}
\begin{equation}
A=\left[\frac{3(1-\nu^2)}{4E}rF\right]^{1/3}{\rm \ ,}
\label{radius}
\end{equation}
$E$ being the Young's modulus, $\nu$ the Poisson's ratio and $r$ the radius of the beads. For stainless steel beads used in the chain [see bead properties in Table\ (\ref{properties})], when $F$ ranges from 10 to 500 N, Eq.\ (\ref{hertz}) leads to a range of deformations $\delta$ between two beads from $2$ to $20$ $\mu$m, and Eq.\ (\ref{radius}) to a range of the contact radii $A$ from $40$ to $200$ $\mu$m.

\begin{figure}[ht]
\resizebox{1\columnwidth}{!}{%
  \includegraphics{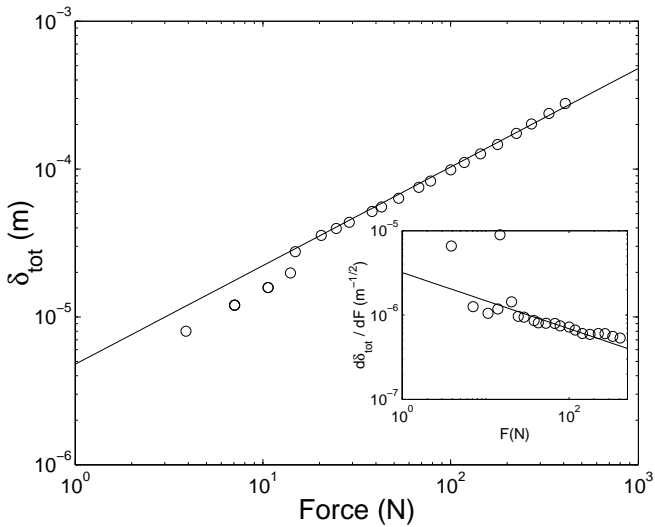}
 }
\caption{Total chain displacement, $\delta_{tot}$, as a function of the applied static force, $F$. (Full line of slope $2/3$). Inset shows $d\delta_{tot}/dF$ vs. $F$ (Full line of slope of $-1/3$). $N=13$.}
\label{fig02}  
\end{figure}

 Figure \ref{fig02} shows the total chain displacement $\delta_{tot}$ as a function of $F$. As expected, there is good agreement with the $F^{2/3}$ Hertz law for our range of $F$. The departure at low $F$ is linked to the fact that $\delta_{tot}$ includes the piston displacement $x_0$, needed to bring all the beads in contact, i.e. $\delta_{tot}=x_0 + 49\delta$. This is indeed shown in the inset of Fig.\ \ref{fig02}, where  $d\delta_{tot}/dF$ is found to scale as $F^{-1/3}$ and to be independent of $x_0$. Using Eq.\ (\ref{hertz}), the intersection of the $F^{-1/3}$ fit with the ordinate axis gives a measurement of the elastic properties of the bead materials through $E/(1-\nu^2)\simeq 10^{11}$ N/m$^2$ in agreement with values extracted from Table\ \ref{properties}. From the sensor documentation, we have checked that the force sensor displacement is 50 times less than the total bead displacement, $\delta_{tot}-x_0$, at $F=500$ N. The mechanical contact of the bead chain is thus very well described by the Hertz law (see also Ref.\ \cite{Coste97,Falcon98}).
 
\section{Electrical behaviour}
\label{results}

 \begin{figure}[ht]
\resizebox{1\columnwidth}{!}{%
  \includegraphics{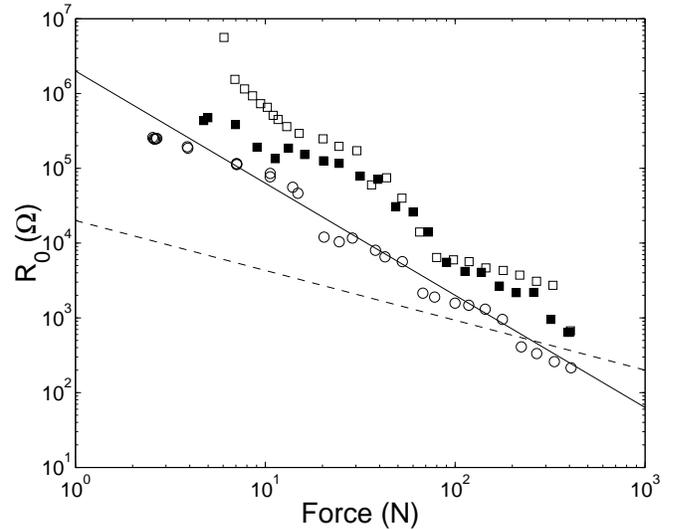}
}
\caption{Electrical resistance, $R_0$, as a function of the applied static force, F, at low imposed voltage $U=10^{-3}$V. ($\circ $) same sample as the one in Fig. \ \ref{fig02}. An other run (see text for details) with increasing $F$ ($\square $), then decreasing $F$ ($\blacksquare$). ($-$) shows $F^{-3/2}$ fit, and ($--$) the $F^{-2/3}$scaling from the Hertz law. $N=13$.}
\label{fig03}  
\end{figure}

 \subsection{Dependence of the resistance on the applied force.}
 \label{forcesection}
Let us denote $R_{0}$ as  the electrical resistance of the bead chain, at low imposed voltage or current. The evolution of $R_{0}$ as a function of the applied force is shown in Fig.\ \ref{fig03}. Experimental points are found to be well fitted by a $F^{-3/2}$ power law (solid line). This measurement is performed simultaneously with the mechanical displacements corresponding to those found in Fig.\ \ref{fig02} which are well described by the Hertz law (see Sect.\ \ref{meca}). Thus, assuming $R_{0} \sim 1/A$ for metallic contact or $R_{0} \sim 1/A^2$ for a slightly oxydized one \cite{Holm00,Bowden39}, Eq.\ (\ref{radius}) leads to an electrical resistance scaling of $F^{-1/3}$ or $F^{-2/3}$, respectively. The unexpected $F^{-3/2}$ scaling observed in Fig.\ \ref{fig03} thus shows that $R_{0}$ does not only depend on $F$ through the radius of contact $A$ but also through the resistivity and thickness of the contaminant and/or oxide film probably present at the interface between metallic surfaces. 

\begin{figure}[ht]
\resizebox{1\columnwidth}{!}{%
  \includegraphics{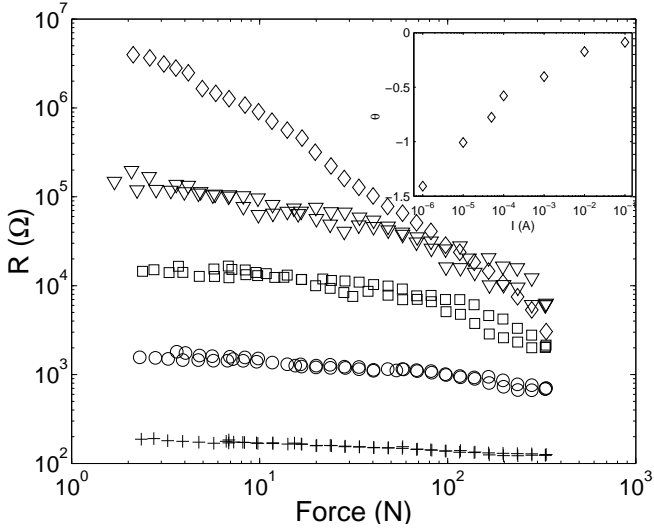}
}
\caption{Electrical resistance, $R$, as a function of $F$, for various current $I$: $10^{-6}$ ($\lozenge $), $10^{-4}$ ($\bigtriangledown $), $10^{-3}$ ($\square $), $10^{-2}$ ($\circ $), $10^{-1}$ ($+$) A. $F$ is increased then decreased. For each curve, $\theta(I)$ exponents are extracted from $R\sim F^{\theta}$ power law fits, and are shown in the inset (semilogx axis). $N=41$.}
\label{fig04n}  
\end{figure}

The $R_0 \sim F^{-3/2}$ scaling is only valid at low current. When $I$ is increased, the $R - F$ law is changed as shown on Fig.\ \ref{fig04n}. For each applied $I$, one can roughly assume a $R\sim F^{\theta}$ power law where $\theta$ is found to be $I$-dependent (see inset of Fig.\ \ref{fig04n}). This complex dependence of $\theta(I)$ comes from the nonlinear characteristics of the system as shown in Sect.\ \ref{characteristics}.

These scaling laws are very robust when repeating our experiments.  After each cycle in force, we roll the beads along the chain axis to have a new and fresh contact between beads for the following cycle.  This is a critical condition to have reproducible measurements.  Indeed, Fig. \ \ref{fig03} shows another force cycle leading to the same low current scaling in $F^{-3/2}$, but shifted vertically by one order of magnitude in resistance. This indicates that, at a given force, $R_{0}$ depends on the film properties at the location where the new contacts have been created. Therefore, $R_{0}$ will subsequently be the control parameter instead of $F$.

Finally, as for the different pressure dependences of the sound velocity observed in a 1D \cite{Coste97} or 2D \cite{Gilles03} granular medium, the electrical resistance scaling ($R_{0} \sim F^{-3/2}$ at low current) should be different for higher dimensions due to the effects of contact disorder and/or percolation. This work is in progress in a 2D hexagonal array of stainless steels beads \cite{Falcon03} or in 3D copper powder samples \cite{Creyssels03}. 

\begin{figure}[ht]
\begin{tabular}{c}
\resizebox{1\columnwidth}{!}{%
 \includegraphics{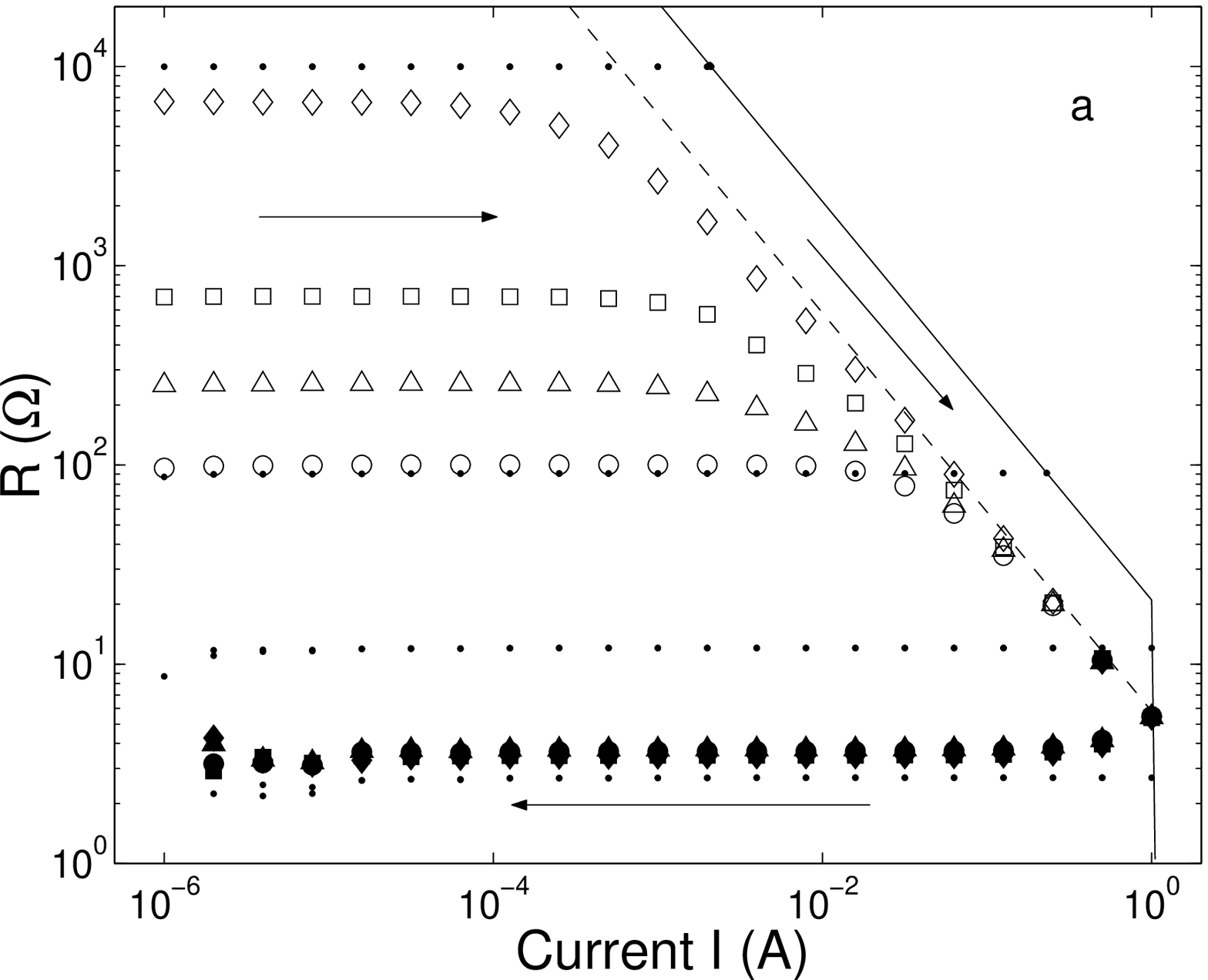}
}\\
\resizebox{1\columnwidth}{!}{%
 \includegraphics{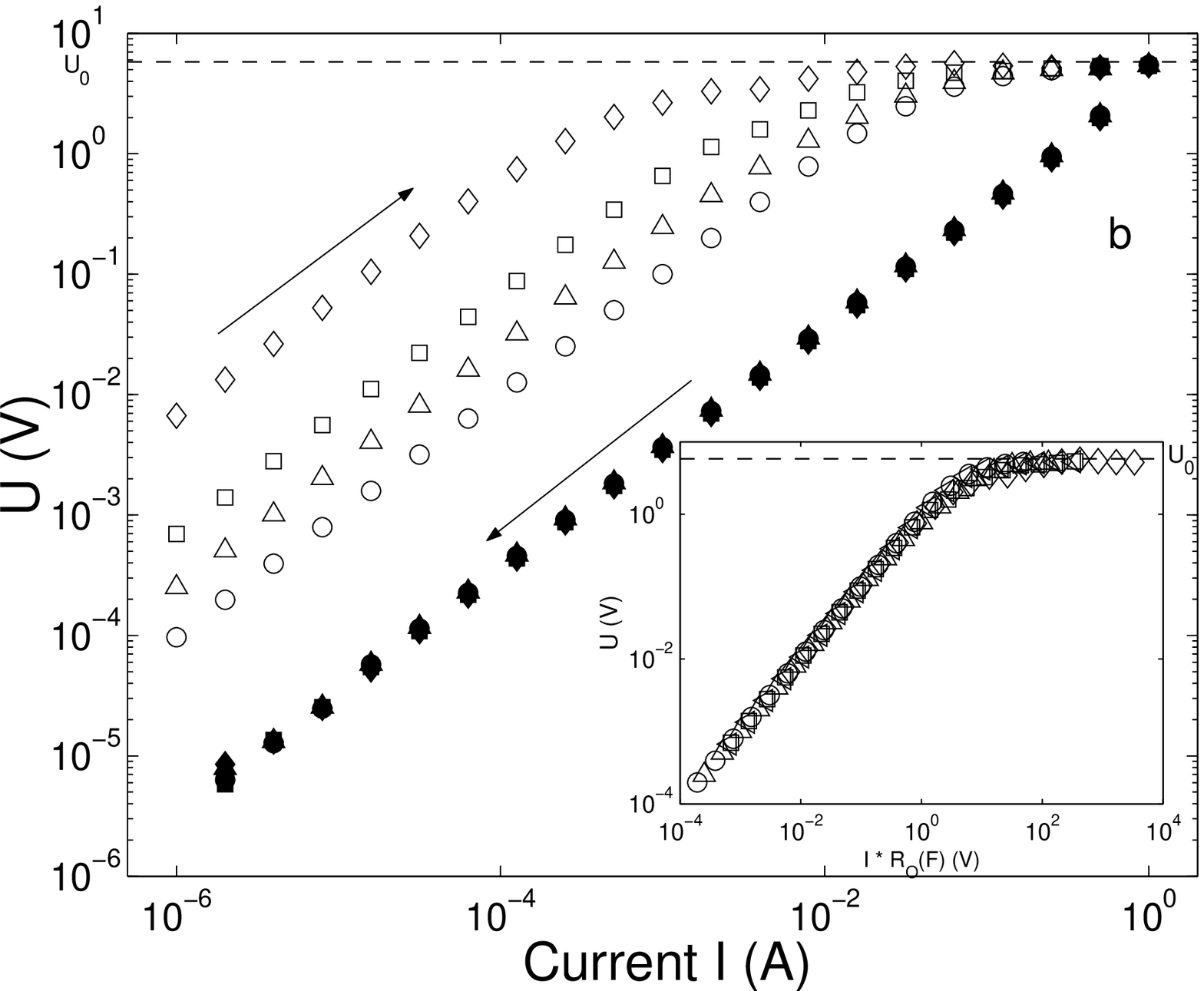}
}
\end{tabular}
\caption{Log-log $R - I$  {\bf (a)} or $U$--$I$  {\bf (b)} characteristics  when increasing the current $I$ (open symbols) from $10^{-6}$ to 1 A, then decreasing $I$ (full symbols) for $N=13$ and various $F=34$ ($\lozenge$), $119$ ($\square $), $305$ ($\bigtriangleup $), $505$ ($\circ $) N. Saturation voltage $U_0=5.8$ $V$ is shown ($--$). Inset shows the $U$--$I$ scaling by the low current resistance $R_0(F)$. Generator maximum compliance values of 21 $V$ and $1.05$ A  ($-$), and measurements of test resistances $R_{test}=10^{4}$, $91$, $12.7$ and $2.7$ $\Omega$ (small $\bullet$-marks) instead of the chain are indicated.}
\label{fig05}  
\end{figure}

\subsection{Nonlinear $U$--$I$ characteristics.}
\label{characteristics}
For various applied forces, Fig.\ \ref{fig05}a shows the typical hysteretical $R$--$I$ characteristics when imposing the current $10^{-6} \leq I \leq 1$ A to the bead chain. At low $I$, the chain resistance $R$ is found to be constant and reversible. As $I$ is increased further (see open symbols), the resistance strongly decreases and reachs a constant bias (see dashed line of slope 1). This will be refered to as the saturation voltage $U_0$. As soon as $U_0$ is reached, the resistance is irreversible upon decreasing the current (see full symbols).  This decrease of the resistance by several orders of magnitude has similar properties as that of the coherer effect with powders \cite{Branly90,Calzecchi84} or with a single contact \cite{Branly02,Branly02bis,Branly99,Branly98,Guthe00,Guthe01}. We have verified these observations are not due to experimental artefacts. The compliance values of the source meter (see solid lines in Fig.\ \ref{fig05}a) are indeed greater than the measured values, and when the chain is replaced by test resistances of known values from 2.7 to $10^{4}$ $\Omega$, measurements (see $\bullet$-marks) lead to the expected results in the full range of currents. As explained previously, after each cycle in the current, the applied force is reduced to zero, and we roll the beads along the chain axis to have new and fresh contact between beads for the next cycle. With this methodology, the resistance drop (coherer or Branly effect) and the saturation voltage are always observed and are very reproducible.

The $U$--$I$ representation of Fig.\ \ref{fig05}a is displayed in Fig.\ \ref{fig05}b. It more easily shows the constant reversible resistance at low current (see open symbols of slope 1), followed by the asymptotical approach to a constant bias value of $U_0$ for larger $I$. When decreasing $I$, it also reveals the irreversible behaviour at another constant resistance (see full symbols of slope 1) having a lower value which depends on the maximum imposed current, but not on the applied force (see Sects.\ \ref{saturation} and \ref{courantinverse}). As mentioned in Sect.\ \ref{forcesection}, the best way to rescale all the $U$--$I$ curves of Fig.\ \ref{fig05}b (performed at various $F$) is not by the force itself but by the resistance at low current, $R_0(F)$. This rescaling is shown in the inset of Fig.\ \ref{fig05}b leading to an impressive collapse on a single master curve. The current-voltage characteristic has thus an ohmic (linear) component which is followed continuously by a nonlinear part saturating for a critical voltage. 

\begin{figure}[ht]
\begin{tabular}{c}
\resizebox{1\columnwidth}{!}{%
\includegraphics{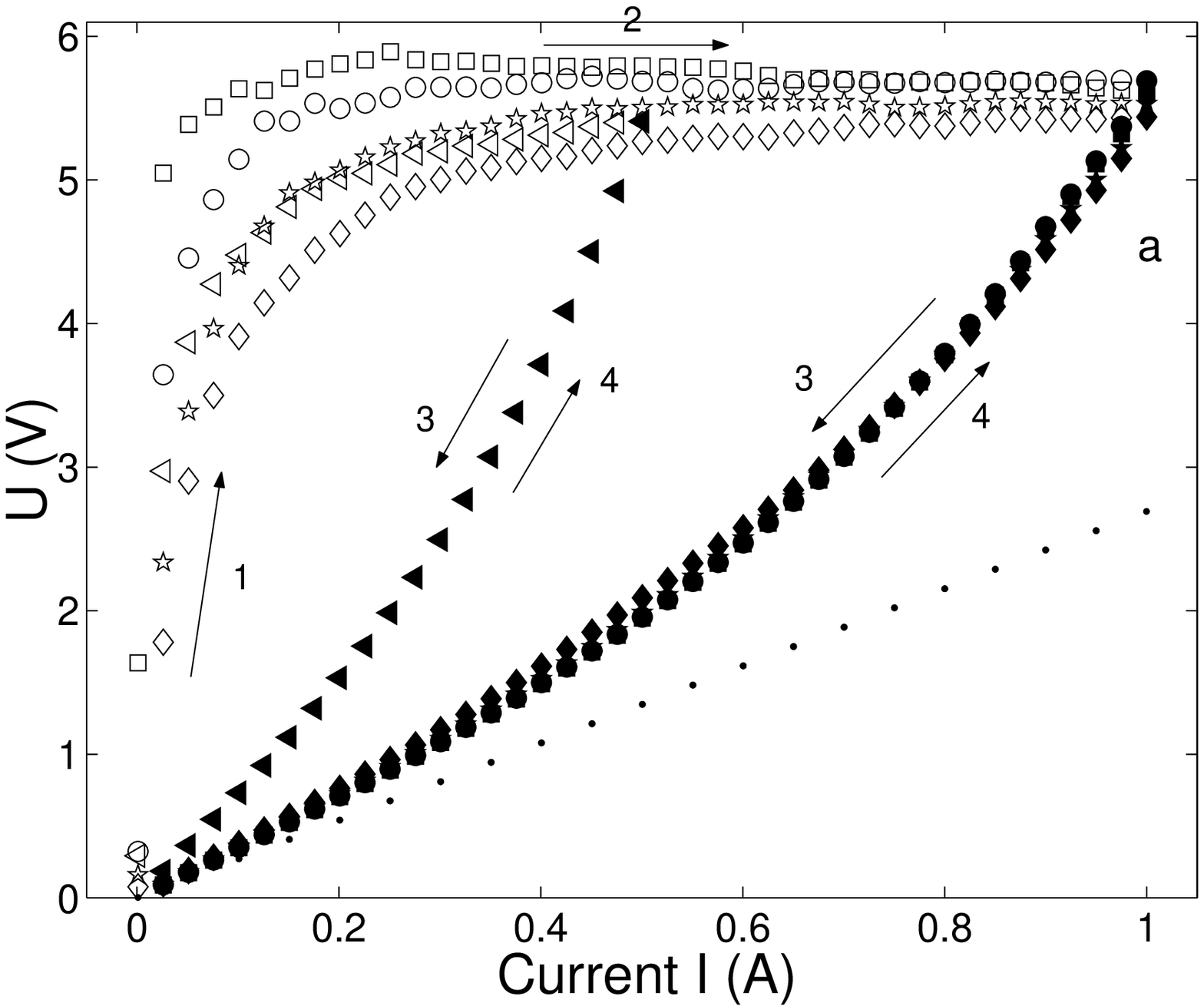}
}\\
\resizebox{1\columnwidth}{!}{%
 \includegraphics{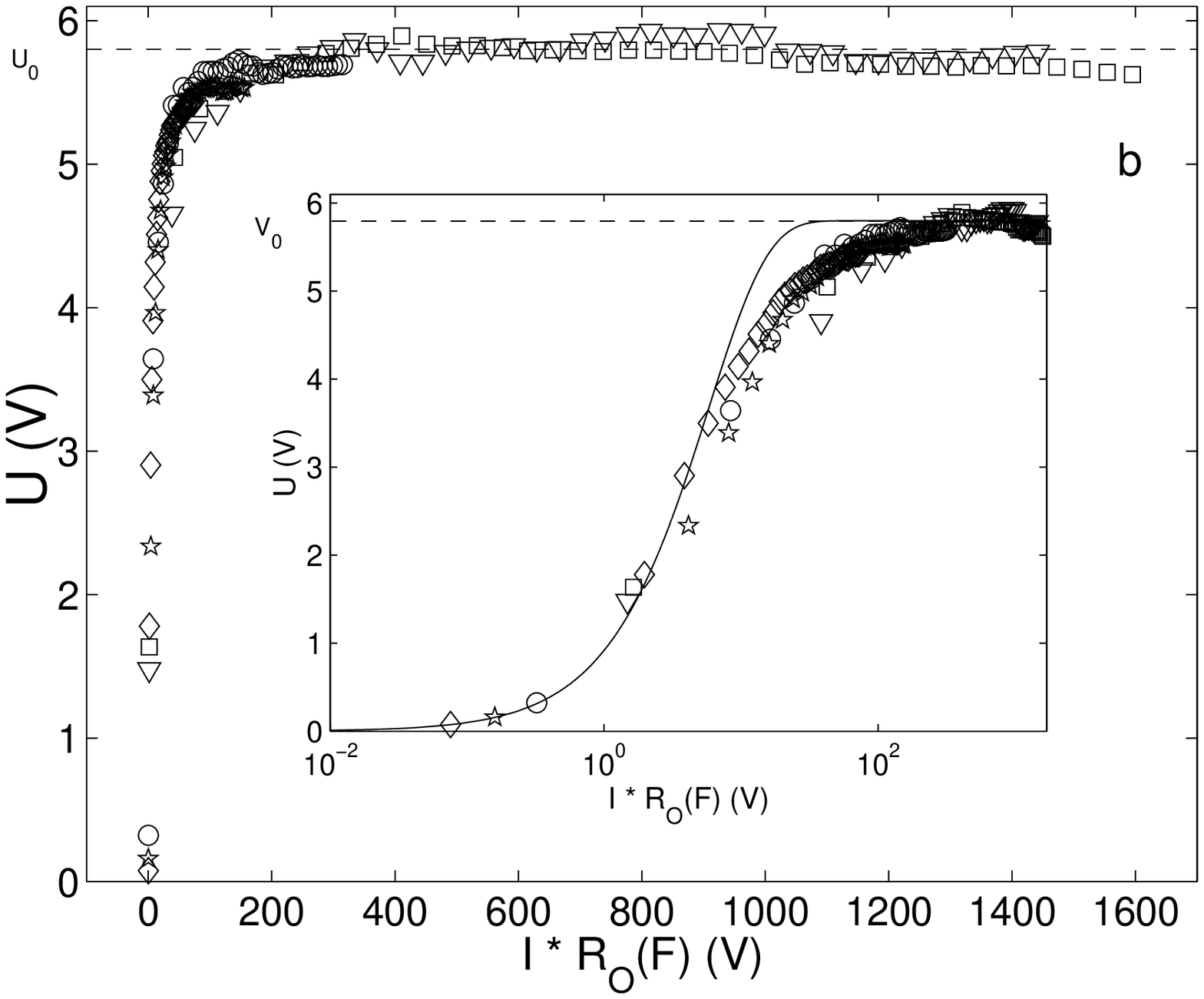}
}
\end{tabular}
\caption{{\bf (a)} $U$--$I$ characteristics (in linear axes) showing the saturation voltage, $U_{0}=5.8$ V, when increasing the current $I$ in the range 1 mA $\leq I \leq I _{max}$ (open symbols), then decreasing $I$ (full symbols) for various $F=32$ ($\square $), 125 ($\circ $), 321 ($\star$) and 505 ($\lozenge$) N with $I _{max}=1$ A, and  for $F=211$ N ($\triangleleft$) with $I _{max}=0.5$ A. Measurement of test resistance $R_{test}=2.7$ $\Omega$ (small $\bullet$-marks) instead of the chain.  {\bf (b)} Same open symbols as (a) rescaled by $R_{0}(F)$ in linear or semilogx axes (inset).  Additional $F=13$ N ($\bigtriangledown $) is shown. Solid line show empirical fit of Guthe \cite{Guthe01}. $N=13$.}
\label{fig06}  
\end{figure}

\subsection{The saturation voltage}
\label{saturation}
In this section, we focus on the saturating regime of the $U$--$I$ behaviour. We performed similar $U$--$I$ studies as described in Sect.\ \ref{characteristics}, but with linear increments of the imposed current.  Fig.\ \ref{fig06}a shows that the characteristic depends on the history of the maximum applied current $I_{max}$.  At low current, $U$--$I$ is reversible and ohmic of resistance $R_0(F)$ (arrow 1). As $I$ is increased, the characteristic follows a constant irreversible line (arrows 2). Then, a decrease from different values of $I_{max}$ leads to different $U$--$I$ trajectories (see full symbols) which are found reversible and non-ohmic (arrows 3 and 4). 

One can show that the saturation voltage $U_0$ depends on the number of beads, $N$, between the electrodes. When varying $N$ from 1 to 41, the saturation voltage per contact $U_{0/c} \equiv U_0 / (N+1)$ is found constant and on the order of 0.4 V per contact as shown in Table\ \ref{Vsat}. These value changes when replacing all stainless steel beads with others of another material, $U_{0/c} \simeq 0.2$ V for bronze beads and $0.3$ V for brass beads (see Table\ \ref{Vsat}). Therefore, $U_{0/c}$ depends slightly on the bead material (see also Ref.\ \cite{Guthe00,Guthe01,Fisch04}), but does not depend on the bead radius \cite{Guthe00} nor on the gas surrounding the beads \cite{Fisch04,Weiss06}. Moreover, the nonlinear saturation bias $U_{0}$ does not depend explicitly on the force $F$. This means that when the previous characteristics obtained for different $F$ are rescaled by $R_0(F)$, all $U$--$I$ curves collapse as shown in Fig.\ \ref{fig06}b on linear or semilog (see inset of Fig.\ \ref{fig06}b) axes. This saturation bias was first observed in 1901 by Guthe \cite{Guthe01} for two beads in contact. He suggested an empirical fit of the form $U=U_0[1-\exp{(-IR_0/U_0)}]$ which does not describe our data (see solid line in the inset of Fig.\ref{fig06}b). If we use a more complex fit, $U=U_0[1-\exp{(-IR_0/U_0)}^{\alpha}]^{\beta}$ with $\alpha\beta=1$ as used in percolation studies \cite{Gupta98}, this leads to a better description but with one adjustable parameter. However, no satisfactory physical description has been proposed for such characteristics and the conduction mechanisms involved. In Sect.\ \ref{interpretation}, we suggest a physical interpretation for $U_{0/c}$ based on an electro-thermal coupling within the microcontacts. Finally, we note that the saturation voltage has not been reported in higher dimensional systems, although these systems exhibit a nonlinear and irreversible $U$--$I$ characteristics (e.g. in 3D polydisperse packing of beads \cite{Dorbolo02} or in 2D metallic packing of pentagons \cite{Dorbolo03}). 

\begin{table}[h]
\caption{Saturation voltage, $U_{0}$, for various bead numbers, $N$, in the chain and for different bead materials. $N_c \equiv (N+1)$ is the number of bead-bead contacts. }
\label{Vsat}  
\begin{tabular}{lclclclcl}
\hline\noalign{\smallskip}
Materials  	& $N_c$ & $U_0$ (V) & $U_0 / N_c $ (V)\\
\hline\noalign{\smallskip}
Stainless Steel & 02 & 0.75  & \bf{0.37}\\
Stainless Steel & 14 & 5.8    &  \bf{0.41}\\
Stainless Steel & 42 &16.5 &  \bf{0.39}\\
Brass 		& 14 & 4.4 & \bf{0.31}\\
Bronze 		& 14 & 3 & \bf{0.21}\\
\hline\noalign{\smallskip}
\end{tabular}
\end{table}

\subsection{Symmetry properties of the $U$--$I$ characteristics}
\label{courantinverse}
Due to contaminants and/or oxide layers probably present on the bead surfaces, a contact between two beads can be described as a Metal/Oxide/Oxide/Metal contact. If the conduction through this contact is ionic or electronic, we expected that the Oxide/Oxide interface has less influence than the Metal/Oxide one. In this case, the conduction via a bridge between metals through the oxide should thus lead to an ionic or electronic accumulation on one side of the contact. When the saturation voltage is reached, it should affect differently the two sides, breaking the original symmetry. This broken symmetry should be observed by an asymmetrical characteristic $U$--$I$, when reversing the applied current to the chain. 

\begin{figure}[ht]
\resizebox{1\columnwidth}{!}{%
 \includegraphics{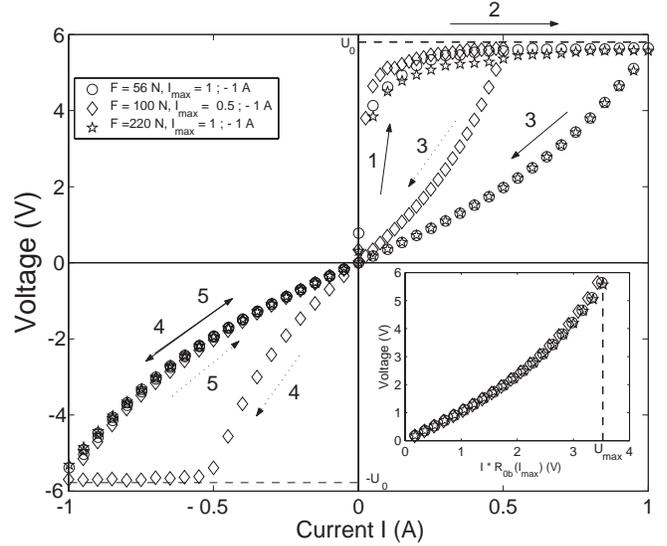}
}
\caption{Symmetrical characteristics $U$--$I$ for various current cycles in the range 1 mA$ \leq I \leq I_{max_{+}}$ and $I_{max_{-}} \leq I \leq $ -1 mA, and for various forces. Inset shows the reversible back trajectories rescaled by $R_{0b}$. $U_{max}\equiv R_{0b} * I_{max}\simeq3.5$ V. $N = 13$. (See text for details).}
\label{fig07}  
\end{figure}

However, when the current is reversed and applied to the chain, Fig.\ \ref{fig07} clearly shows a symmetrical curve $U(I)=-U(-I)$. At low applied current, $U$--$I$ is reversible and ohmic (solid arrow 1), then it nonlinearly reaches the irreversible saturation regime (solid arrow 2) for increasing $I$ up to $I_{max_{+}}\ $ $=1$ A, and finally follows a nonlinear and reversible back trajectory (solid arrow 3) when $I$ is decreased to 1 mA. When reversing the current up to $I_{max_{-}}=-1$ A, the characteristic follows this reversible non-ohmic line symmetrically (solid arrows 4 and 5). We can thus conclude that the important interface is the Oxide/Oxide interface in this case.

We now repeat the experiment (see $\lozenge$-marks) up to a different $I_{max_{+}}=0.5$ A. It leads to a different back trajectory (dashed arrow 3) which is again symmetrical when reversing the applied current up to $I_{max_{-}}=-0.5$ A  (dashed arrow 4). When the current is further decreased up to -1 A, the characteristic symmetrically reaches the saturation bias $-U_0$ (see $\lozenge$-marks), before joining the previous reversible non-ohmic line (dashed arrow 5), when the current is increased from -1 A to 1 mA. Thus, the back trajectory of this symmetrical loop is driven by $|I_{max}|$.  To check this, let us first define $R_{0b}$ as the electrical resistance of the chain, at low decreasing current, that is, the slope of the back trajectory in Fig.\ \ref{fig07}.  When repeating this experiment up to different values of $I_{max}$ and for various applied forces $F$, one can show that $R_{0b}$ does not depend on $F$, but only on $I_{max}$ such as $R_{0b}*I_{max} \equiv U_{max}$ is constant. It is indeed shown in the inset of Fig.\ \ref{fig07} where the reversible back parts of Fig.\ \ref{fig07} are rescaled by $R_{0b}$, and follow the same back trajectory.

\section{Interpretation}
\label{interpretation}

\subsection{Qualitative interpretation}
\label{interpret} 
\begin{figure}[ht]
\resizebox{1\columnwidth}{!}{%
 \includegraphics{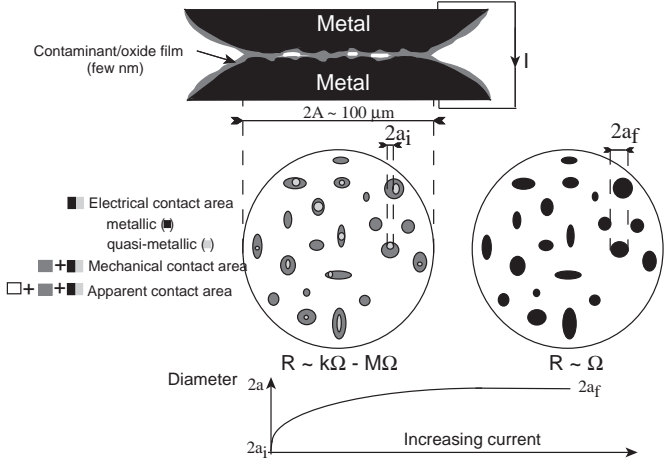}
}
\caption{Schematic view of the electrical contact building through microcontacts by transformation of the poorly conductive contaminant/oxide film. At low $I$, the electrical contact is mostly driven by a complex conduction mechanism through this film via conductive channels (of areas increasing with $I$). At high enough $I$, an electro-thermal coupling generates a soldering of the microcontacts leading to efficient conductive metallic bridges (of constant areas).}
\label{fig08}  
\end{figure}

Assume a mechanical contact between two metallic spheres covered by a thin contaminant film ($\sim$ few nm). The interface generally consists of a dilute set of microcontacts due to the roughness of the bead surface at a specific scale \cite{Holm00}. The mean radius, $a$, of these microcontacts is of the order of magnitude of the bead roughness $\sim$ 0.1 $\mu$m, which is much smaller than the apparent Hertz contact radius $A\sim 100$ $\mu$m. Figure\ \ref{fig08} schematically shows the building of the electrical contact by transformation of this poorly conductive film. At low applied current, the high value of the contact resistance (k$\Omega$ -- M$\Omega$) probably comes from a complex conduction path \cite{DaCosta00} found by the electrons through the film within a very small size ($\ll 0.1\ \mu$m) of each microcontact (see lightly grey zones in Fig.\ \ref{fig08}). The electron flow then damages the film, and leads to a ``conductive channel'': the crowding of the current lines within these microcontacts generates a thermal gradient in their vicinity, if significant Joule heat is produced. The mean radius of microcontacts then strongly increases by several orders of magnitude (e.g., from $a_i\ll 0.1\ \mu$m to $a_f\sim 10\ \mu$m), and thus enhances their conduction (see Fig.\ \ref{fig08}). This corresponds to a nonlinear behaviour (arrow 1 until 2 in Fig.\ \ref{fig07}). At high enough current, this electro-thermal process can reach the local soldering of the microcontacts (arrow 2 in Fig.\ \ref{fig07}); the film is thus ``piercing'' in a few areas where purely metallic contacts (few $\Omega$) are created (see black zones in Fig.\ \ref{fig08}).  [Note that the current-conductive channels (bridges) are rather a mixture of metal with the film material rather than a pure metal. It is probable that the coherer action results in only one bridge -- the contact resistance is lowered so much that puncturing at other points is prevented]. The $U$--$I$ characteristic is then reversible when decreasing and then increasing $I$ (arrow 3 in Fig.\ \ref{fig07}). The reason is that the microcontacts have been soldered, and therefore their final size $a_f$ does not vary any more with $I < I_{max}$. The $U$--$I$ back trajectory then depends only on the temperature reached in the metallic bridge through its parameters (electrical and thermal conductivities), and no longer on its size as previously.
 
\subsection{Quantitative interpretation}

To check quantitatively the interpretation in \ref{interpret}, we shall first recall the voltage - temperature $U$--$T$ relation, and we shall see that this electro-thermal coupling is the simplest way to interpret quantitatively the $U$--$I$ back trajectory (arrows 3 in Fig.\ \ref{fig07}). Indeed, the relationship between the voltage-drop across the contact $U$, the current $I$ and the microcontact radius $a$ is strongly modified compared to the classical constricted case, $U/I=\rho_{el}/2a$, derived without taking into account the significant heat production within the microcontact. 

Assume a microcontact between two clean metallic conductors (thermally insulated at uniform temperature $T_0$, with no contaminant or tarnish film at the contact). Such a clean microcontact is generally called a ``spot". If an electrical current flowing through this spot is enough to produce Joule heating, then a steady-state temperature distribution is quickly reached ($\sim \mu$s) in the contact vicinity.  The maximum temperature $T_m$ is located at the contact, and is linked to the voltage-drop $U$ by the Kohlrausch's equation \cite{Holm00,Greenwood58,Timsit99}
\begin{equation}
U^2=8\int_{T_0}^{T_m}\lambda \rho_{el}\ dT  {\rm \ ,}
\label{Kohlrausch}
\end{equation}
where, $\rho_{el}$ is the electrical resistivity and $\lambda$ the thermal conductivity, both being dependent on the temperature $T$. However, for many conductors, the Wiedemann-Franz law states that \cite{Holm00,Timsit99,Fechant96}
\begin{equation}
\lambda \rho_{el} = LT\ {\rm ,} 
\label{WFlaw}
\end{equation}
where $L=\pi^2 k^2/(3e^2)=2.45\ 10^{-8}$ V$^2$/K$^2$ is the Lorentz constant, $k$ the Boltzmann's constant, and $e$ the electron charge. This is a consequence that the electron scattering time contributes to both the electrical conductivity and the heat conductivity. Combining Eqs.\ (\ref{Kohlrausch}) and (\ref{WFlaw}) yields for the local heating;
\begin{equation}
T_m^2 - T_0^2 = \frac{U^2}{4L}  {\rm \ .}
\label{WF}
\end{equation}
This relationship shows that the maximum temperature $T_m$ reached at the contact is independent of the contact geometry and of the materials in contact! This is a consequence that both the electrical and thermal conductivities are related to the conduction electrons through Eqs.\ (\ref{WFlaw}). However, the temperature distribution within the bridge depends on the geometry \cite{Fechant96}. A voltage near $0.4$ V across a constriction thus leads  from Eq.\ (\ref{WF}) to a contact temperature near 1050$^{\rm o}$C for a bulk temperature $T_0=20^{\rm o}$C. This means that $U\simeq 0.3-0.4$ V leads to contact temperatures that exceed the softening or/and the melting point of most conductive materials \cite{Fechant96}. Efficient conductive metallic bridges (or ``hot spots'') are therefore created by microsoldering. Moreover, Eq.\ (\ref{WF}) shows that the parameter determining the spot temperature is the voltage-drop across the contact, not the magnitude of the current: this explains why the experimental saturation voltage $U_{0/c}$ is the relevant parameter in Sect.\ \ref{results}. In addition, when $U$ approaches $U_{0/c}$, the local heating of microcontacts is enough, from Eq.\ (\ref{WF}), to soft them (mainly at their peripheries \cite{Fechant96}). Then, their contact areas increase thus leading to a decrease of local resistances, and thus stabilizing the voltage, the contact temperatures and the contact areas, since the current is fixed. The phenomenon is therefore self-regulated in voltage and temperature. 
%{\em decrease of contact temperatures, the contact areas then decrease, and thus the local resistances and temperatures increase}. 

Let us now specify the temperature dependance for the thermal and electrical conductivity in the case of an alloy or a pure metal. For an alloy, some defects are present in the bulk metal, and contribute to the electrical conductivity (but have no influence on the Eq.\ (\ref{WFlaw}) \cite{Holm00}). The elastic scattering of the conductive electrons with the metal phonons and with the defects is random, and thus the corresponding scattering frequencies add up. This leads to the Mathiessen's rule: the total resistivity is the sum of a temperature dependant resistivity due to the scattering with phonons and a residual resistance at zero temperature due to defects such as 
\begin{equation}
\rho_{el}(T)\equiv \alpha(T+\beta T_0)
\label{alloy}
\end{equation}
where $\alpha=6.98\ 10^{-10}$ $\Omega$.m/K and $\beta=2.46$ are extracted from the stainless steel resistivity in Table~\ref{properties}, and $T_0=293$ K is the room temperature. This defines an effective temperature linked to the defects $T_{def}\equiv \beta T_0$. Note that for pure metals, $\alpha=\rho_0/T_0$, $\beta=0$ (since only the ``phonon resistivity'' contributes), and thus from Eq.\ (\ref{WFlaw}), $\lambda(T)= \lambda_0=LT_0/\rho_0$, where $\rho_0$ and $\lambda_0$ are the electrical resistivity and thermal conductivity of the pure metal at $T_0$. 

One can analytically solve the electro-thermal problem for the general case of an alloy, i.e. with $\rho_{el}(T)$ and $ \lambda(T)$ as in Eqs.\ (\ref{WFlaw}) and (\ref{alloy}), as shown in the appendix:
\begin{itemize}
\item the isothermal temperature $T_m$ at the contact surface as a function of the voltage $U$ is
\begin{equation}
T_m=\sqrt{T_0^2+\frac{U^2}{4LN_c^2}}\ {\rm ,}  
\label{solT}
\end{equation}

\item the normalized current $IR_{0b}$ through the contacts only depends on this temperature $T_m$ (i.e. on $U$) such as
\begin{equation}
IR_{0b}=2T_0N_c\sqrt{L}(1+\beta)\int_0^{\theta_0}\frac{\cos{\theta}}{\beta \cos{\theta_0}+\cos{\theta}}d\theta\ {\rm ,}
\label{solIalloy}
\end{equation} 
where $\theta_0 \equiv \arccos{(T_0/T_m)}$.
\end{itemize}
We remind the reader that $T_m$ does not depend on the material properties, or the microcontact geometry, but only on the room temperature, $T_0$, the number of bead-bead contacts in the chain, $N_c \equiv N+1$, and the Lorentz constant $L$. $IR_{0b}$ has an additional parameter $\beta$ related to the defects in the material. For pure metals, i.e with $\alpha=\rho_0/T_0$ and $\beta=0$ in Eq.\ (\ref{alloy}), Eq.\ (\ref{solIalloy}) simplifies to the well-known explicit expression with no adjustable parameter \cite{Holm00,Greenwood58}
\begin{equation}
IR_{0b}=2N_cT_0\sqrt{L\arctan[T^*_m(2+T^*_m)]}\ {\rm ,}
\label{solImetal}
\end{equation}
where $T^*_m\equiv (T_m- T_0)/T_0$.

\begin{figure}[ht]
\resizebox{1\columnwidth}{!}{%
 \includegraphics{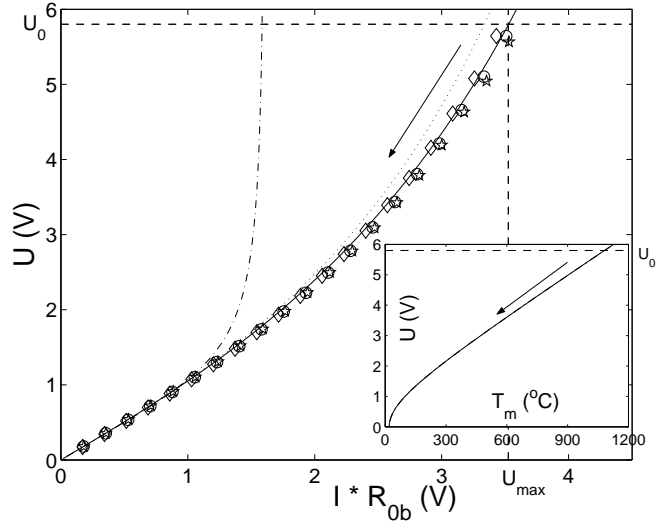}
}
\caption{Comparison between experimental $U$--$I$ back trajectories of Fig.\ \ref{fig07} (symbols), and theoretical curves for an alloy [Eq.\ (\ref{solIalloy})] with stainless steel properties [$\beta=3$ ($-$) or $\beta=2.46$ ($\cdots$)], and for a pure metal ($-.-$) [Eq.\ (\ref{solImetal})]. Inset shows the theoretical maximum temperature, $T_m$  [Eq.\ (\ref{solT})], reached in one contact when the chain is submitted to a voltage $U$. $N=13$.}
\label{fig09}  
\end{figure}

The normalized $U$--$I$ back trajectory (i.e., $IR_{0b}$ as a function of $U$) is displayed in Fig.\ \ref{fig09}. Here, the experimental results of Fig.\ \ref{fig07} are compared with the theoretical solutions for an alloy [Eq.\ (\ref{solIalloy})] calculated with AISI 304 stainless steel properties, and for a pure metal [Eq.\ (\ref{solImetal})]. A very good agreement is shown between the experimental results and the electro-thermal theory, notably for the alloy case. Qualitatively, the alloy solution has a better curvature than the pure metal one. The agreement is quantitatively excellent when choosing $\beta=3$ instead of $2.46$ (the $\beta$ value for AISI 304 stainless steel), since the $\beta$ value for the bead material (AISI 420 stainless steel) is unknown, but should be close. This gives a measurement of the effective temperature due to the defects $T_{def}=3T_0$. During this experimental back trajectory, the equilibrium temperature, $T_m$, on a microcontact is also deduced from Eq.\ (\ref{solT}) with no adjustable parameter (see inset of Fig.\ \ref{fig09}). Therefore, when the saturation voltage is reached ($U_0=5.8$ V), $T_m$ is close to 1050$^{\rm o}$C which is enough to soften or to melt the microcontacts. Our implicit measurement of the temperature is equivalent to use a resistive thermometer. When very high voltage (more than 500 V) is applied to a monolayer of aluminium beads, direct visualization with an infrared camera has been performed by Vandembroucq et al. \cite{Vandembroucq97}. 
 
\section{Conclusion}
We have reported the observation of the electrical transport within a chain of oxidized metallic beads under applied static force.  A transition from an insulating to a conductive state is observed as the applied current is increased. The $U$--$I$ characteristics are nonlinear, hysteretic, and saturate to a low voltage per contact ($\simeq 0.4$ V).  Electrical phenomena in granular materials related to this conduction transition such as the ``Branly effect'' were previously interpreted in many different ways but without a clear demonstration. Here, we have shown that this transition, triggered by the saturation voltage, comes from an electro-thermal coupling in the vicinity of the microcontacts between each bead.  The current flowing through these spots generates local heating which leads to an increase of their contact areas, and thus enhances their conduction. This current-induced temperature rise (up to 1050$^{\rm o}$C) results in the microsoldering of contacts (even for so low voltage as 0.4 V). Based on this self-regulated temperature mechanism, an analytical expression for the nonlinear $U$--$I$ back trajectory is derived, and is found in very good agreement with the data. It also allows the determination of the microcontact temperature all through this reverse trajectory, with no adjustable parameter. Finally, the stress dependence of the resistance is strongly found non-hertzian underlying a contribution due to the surface films. 

\begin{acknowledgement}
We thank D.~Bouraya for the realization of the experimental setup, and G.~Kamarinos for sending us Ref. \cite{Kamarinos75,Kamarinos90}. L.~K.~J.~Vandamme and E.~Guyon are grateful for the fruitful discussions.
\end{acknowledgement}

\appendix\section{Appendix}
Assume a single plane contact (of any shape) between two identical conductors (of large dimensions compared to the contact ones) submitted to a  constant current $I$. The electrical power dissipated by the Joule effect is assumed totally drained off by thermal conduction in the conductors. This thermal equilibrium and Ohm's law lead to the potential $\varphi$ at the isotherm $T$ in the contact vicinity \cite{Holm00,Greenwood58,Fechant96},
\begin{equation}
\varphi^2(T)=2\int_T^{T_m}\rho_{el}(T')\lambda(T')dT'\ {\rm ,}
\label{Phi-T}
\end{equation}
where $T_m$ is the maximum temperature occurring in the contact plane, $\lambda$ the thermal conductivity and $\rho_{el}$ the electrical resistivity of the conductor. Denote by $R_{0b}$, the ``cold'' contact resistance presented to a current low en\-ough not to cause any appreciable rise in the temperature at the contact (the conductor bulk being at the room temperature $T_0$). The relation between the current flowing through the contact and the maximum temperature produced is then \cite{Greenwood58} 
\begin{equation}
IR_{0b}=2\rho_{el}(T_0)\int_{T_0}^{T_m}\frac{\lambda(T)}{\varphi(T)}dT \ {\rm .} 
\label{Ia}
\end{equation}
Note that the dependence on temperature of the right-hand side of Eq.\ (\ref{Ia}) arises solely from the presence of material parameters, and that only $R_{0b}$ depends on the contact geometry. Solving this equation for the general case of $\lambda (T)$ and $\rho_{el}(T)$ such as in Eqs.\ (\ref{WFlaw}) and (\ref{alloy}). Substituting Eqs.\ (\ref{WFlaw}) and (\ref{alloy}) in the so-called ``$\varphi$--$T$'' relation [the Kohlrausch's equation\ (\ref{Phi-T})] leads to Eq.\ (\ref{solT}) for $N_c$ contacts in series, since $\varphi(T_0)\equiv U/2$. Substituting Eqs.\ (\ref{WFlaw}) and (\ref{alloy}) in Eq.\ (\ref{Ia}) yields to
\begin{equation}
IR_{0b}=2\sqrt{L}T_0(1+\beta)\int_{T_0}^{T_m}\frac{\frac{T/T_0}{\beta+T/T_0}}{T_m\sqrt{1-\left(T/T_m\right)^2}}dT\ {\rm .}
\label{int}
\end{equation}
Making the change of variable $\theta\equiv \arccos{(T/T_m)}$, Eq.\ (\ref{int}) reduces to
\begin{equation}
IR_{0b}=2\sqrt{L}T_0(1+\beta)\int_{0}^{\theta_0}\frac{\cos{\theta}}{\cos{\theta}+\beta\cos{\theta_0}}d\theta\ {\rm ,}
 \label{eqq}
\end{equation}
with $\theta_0 \equiv \arccos{(T_0/T_m)}$. For $N_c$ contacts in series, Eq.\ (\ref{eqq}) leads to Eq.\ (\ref{solIalloy}).

\bibliographystyle{unsrt}
\bibliography{1DconducV3}

\end{document}